\input{aipcheck}

\documentclass[
    ,final            
  ]
  {aipproc}

\layoutstyle{8x11single}
\usepackage{float}
\usepackage{graphicx}
\usepackage{subfig}

\begin{document}

\title{Phase controlling current reversals in a  chaotic ratchet transport}

\classification{05.45.-a, 87.15.hj, 05.60.Cd}
								
\keywords      {Directed transport, ratchet transport, Current reversal, chaos, bifurcation diagrams, current reversal diagrams}

\author{Bruno S. Dandogbessi and Omololu Akin-Ojo}{
address={Theoretical Physics Department,  African University of Science and Technology, Km 10 Airport Road, Galadimawa, Abuja, Nigeria }
}

\author{Anatole Kenfack}{
address={Physikalische und Theoretische Chemie, Institut f{\"u}r Chemie und Biochemie, Freie Universit{\"a}t Berlin, Takustr.3, DE-14195 Berlin, Germany}
}

\begin{abstract}
We consider a deterministic chaotic ratchet model for which the driving force 
is designed to allow the rectification of current as well as the control
of chaos of the system. Besides the amplitude of the symmetric driving force which is 
often used in this framework as control parameter, a phase has been newly included here. 
Exploring this phase, responsible of the asymmetry of the driven force, a number of 
interesting departures have been revealed. Remarkably, it becomes possible to drive 
the system into one of the following regime: the state of zero transport, the state of directed transport and most importantly the state of reverse transport (current reversal).
To have a full control of the system, a current reversal diagram has been 
computed thereby clearly showing the entire transport spectrum which is expected to be 
of interest for possible experiments in this model.  
\end{abstract}

\maketitle


\section{Introduction}

In the past decades, the directed transport induced by symmetry breaking under forces of zero mean 
has been the topic of great interest in many fields. This idea goes  back to the works of Smoluchowski and Feynman~\cite{smoluchowski1912} 
and includes the study of the so-called ratchets, initially motivated by the mechanism underlying 
the functioning of molecular motors in biological systems~\cite{haenggi1996}. Soon after, this ratchet 
mechanism found its application in many domains of classical and quite recently of quantum physics; see~\cite{haenggi2009} 
for a recent and comprehensive review. 

Among the many kinds of ratchets so far 
explored,  an important class refers to classical deterministic ratchets in which the dynamics
does not have any randomness or stochastic elements. In these deterministic models for which the role of noise was successfully replaced by chaos due to the inertial term~\cite{jung1996}, a number of interesting phenomena has 
been revealed.  Along these lines, current reversal turns out to be particularly fascinating and surprising as it 
happens to be very counterintuitive. The 
interest to explain how this arises and its physical origin has been continuously 
increasing, thereby leading to interesting theoretical and experimental 
contributions(~\cite{mateos2000, barbi2000, reimann1997, kenfack2007,arzola2011}, to name a few). Not to mention, Hamiltonian ratchets have recently
seen a breakthrough in the ratchet community~\cite{schanz2001}. Theses systems, for which the 
coherence is fully preserves, owe their 
merit to the first experimental  realization of the quantum ratchet potential~\cite{ritt2006,salger2007} which has considerably inspired theoretical as well as
experimental works. Remarkably in this context, current reversals have also been found with atoms~\cite{lundh2005, kenfack2008}, and directed transport of
atoms has quite recently been experimentally achieved~\cite{salger2009}. 

More specifically, in a chaotic 1D classical deterministic ac-driven ratchet model, Mateos predicted 
the occurrence of the current reversal and has associated it with bifurcation from 
chaos to regular regimes~\cite{mateos2000}. Because such a chaotic system can exhibit coexistence 
of attractors (chaotic or not) likely to transport independently either to the left or to the right, statistical
calculations must be taken into account, in contradiction to a single particle calculation which 
is erroneous~\cite{barbi2000}. This misconception was recently clarified by Kenfack et al.~\cite{kenfack2007} and has led to a new conjecture: Abrupt changes in the current are correlated with 
bifurcations, but do not lead always to current reversals. The detailed mechanism of such a surprising effect
is still an open question; In a quantum system, this effect has been associated to tunneling~\cite{reimann1997}. 

In this work, we address the transport properties of a particle in a chaotic deterministic ratchet model, 
the same as in Ref.~\cite{mateos2000}, but with a novelty: the driving force which is asymmetric 
is designed with two parameters likely to control the current reversal on the systems. We show that we are able to control
a particle transport which can resume a non trivial behavior as function two important parameters, namely the amplitude and the phase of the driving force. Furthermore we come up with a rather completely two parameters current reversal diagram likely to indicate regions of current reversals (reverse transport) and regions of purely directed transport.

The outline of the paper is as follows. The description of our model is presented in section II, while section III provides few details on a particle transport quantities of our interest. Section IV shows the results and comments of our numerical experiments and section IV concludes the paper. 
 
\section{The Model}
\begin{figure}[hc]
\includegraphics[width=0.55\linewidth,height=.35\textheight]{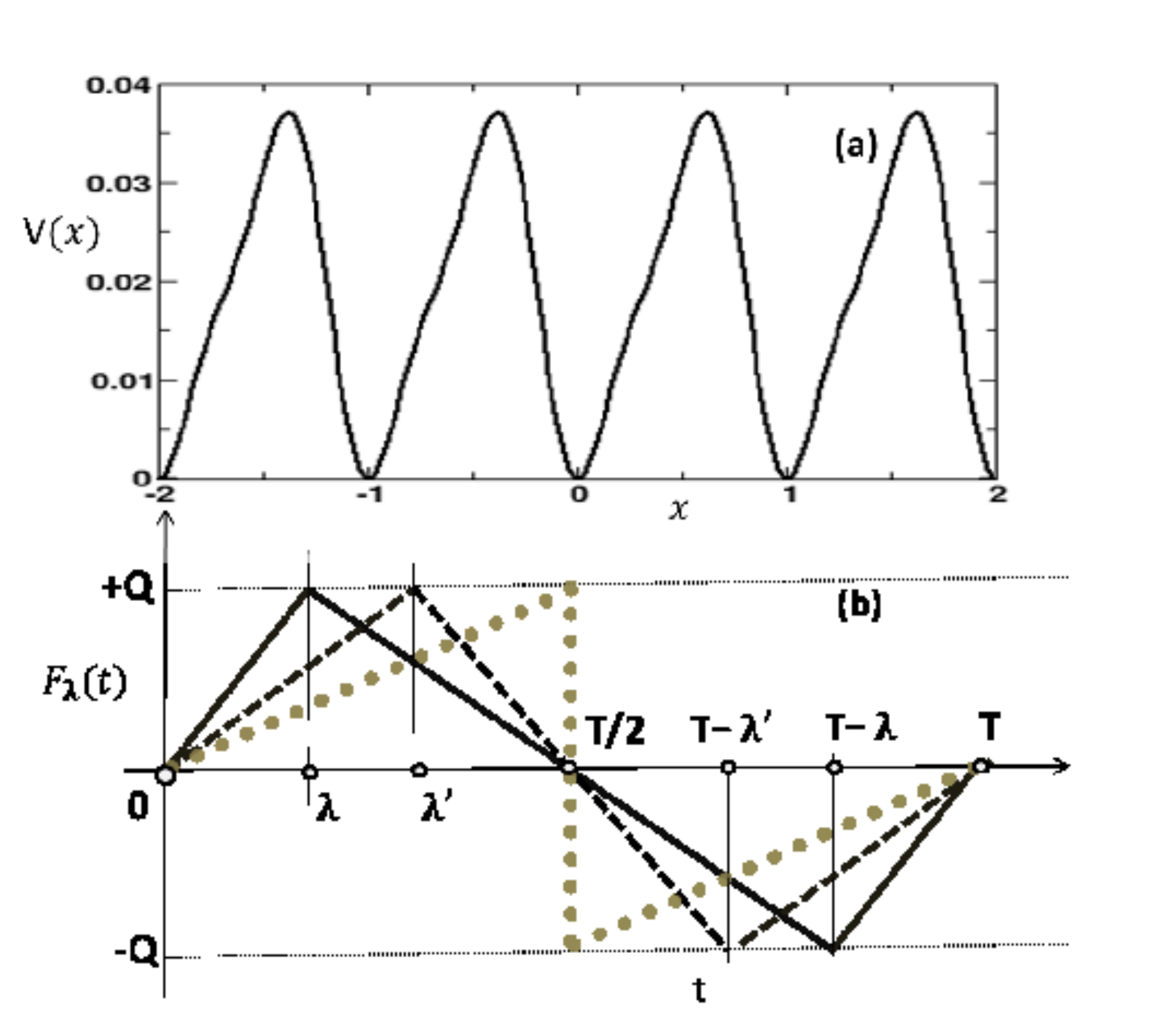}
   \caption{(a) The dimensionless ratchet potential $v(x)$ as function of the position $x$; (b) the dimensionless driving force $F_\lambda(t)$ as function of the time $t$ for given period $T$, amplitude $Q$ and asymmetry parameter $\lambda$. Note that $\lambda$ $\in (0,T/2]$ and $v(x)$ is plotted for arbitrary $\lambda$ (solid), $\lambda'$ (dashed) and the limiting case $\lambda=T/2$ (dotted).}
	\label{fig1}
\end{figure}

The system under consideration is a one-dimensional problem for a particle experiencing a ratchet like spatial potential and under the influence of an external time dependent asymmetric driving force whose average is zero. The system is deterministic as we do not account for any type of noise. The associated equation of motion reads:
\begin{eqnarray}
m\ddot{x}+\gamma\dot{x}+\frac{dV(x)}{dx}=F_\lambda(t)
\label{eom1}
\end{eqnarray}
where $m$ is the mass of the particle, $\gamma$ the friction coefficient, $V(x)$ the external ratchet potential, $F_\lambda$ the time dependent external force. The ratchet potential has the following expression:
\begin{eqnarray}
V(x)=V_1-V_0\sin(\frac{2\pi}{L}(x-x_0))-\frac{V_0}{4}\sin(\frac{4\pi}{L}(x-x_0))
\label{eq2}
\end{eqnarray}
where $L$ is the periodicity of the potential, $V_1$ is an arbitrary constant, $V_0$ the potential amplitude. This potential is shifted by $x_0$ in order to bring the potential minimum to the origin. The essential of the dynamics of our model is driven by the time dependent external force of the following form:
\begin{eqnarray}
		F_\lambda(t) = \left\{ 
  \begin{array}{l l}
    \frac{t-nT/2}{\lambda} Q_0 & \quad \textrm{if $nT/2\leq t\leq \lambda+nT/2$ }\\
    \frac{t-(n+1)T/2}{\lambda - T/2}Q_0 & \quad \textrm{if $\lambda +nT/2\leq t\leq (n+2)T/2-\lambda$}\\
		\frac{t-(n+2)T/2}{\lambda}Q_0 & \quad \textrm{if $(n+2)T/2-\lambda \leq t \leq (n+2)T/2$}
  \end{array}\right .
	\label{dforce}
	\end{eqnarray}
where $n$ is an integer that counts the number of period $T$ evolved in time. This force is thus characterized by the amplitude $Q_0$ and the parameter $\lambda$. Here $\lambda$ regulates the asymmetry of the force and is thus conveniently called asymmetry parameter. Next we use the following dimensionless units: $x'=x/L$, $x'_0=x_0/L$, $t'=\omega_0t$, $T'=\omega_0T$, $\lambda'=\omega_0\lambda$, $b=\gamma/m\omega_0$, and $Q=Q_0/m\omega^2L$; where $\delta=\sin(2\pi|x_0|)+1/4\sin(4\pi|x_0|)$ and $\omega^2=4\pi^2V_0\delta/mL^2$. Here $w_0$ stands for the frequency of the linear small oscillations around the potential minima and $T_0=2\pi/w_0$ the corresponding natural period which is our time scale. It turns out that the dimensionless equation of motion, for which the variables have been renamed for simplicity, takes the form:
\begin{eqnarray}
\ddot{x}+b\dot{x}+\frac{dV(x)}{dx}=F_\lambda(t)
\label{eq3}
\end{eqnarray}
since the only change on the driving force is the amplitude $Q_0$ which has been rescaled to $Q$, while the potential potential $v(x)=C-(\sin(2\pi(x-x_0))+1/4\sin(4\pi(x-x_0)))/4\pi^2\delta$, where $C=V_1/m\omega^2L^2$ is merely a constant. Fig.~\ref{fig1} displays within one period the spatial ratchet potential $v(x)$ (Fig.~\ref{fig1}-a) and the asymmetric driving and periodic force $F_\lambda(t)$ (Fig.~\ref{fig1}-b) plotted for the first period as well, that is for $n=0$. The asymmetry of the driving force is indeed realised by simply varying the asymmetry parameter $\lambda$. For a given period $T$, $\lambda$ takes values in $(0,T/2]$ and it is worth pointing out that the asymmetry of the driving force is broken if only if $\lambda$ equals $T/4$. We note that the dimensionality of the variables means that all quantities computed and reported in this work are dimensionless, unless stated otherwise. 

To attempt to achieve our goal, that is to have full control over the particle motion of this system, we endeavour in the subsequent section to compute conveniently suitable quantities likely to uncover the particle transport behavior.


\section{Transport properties of a particle}
\begin{figure}[hc]
\includegraphics[width=0.45\linewidth,height=.25\textheight]{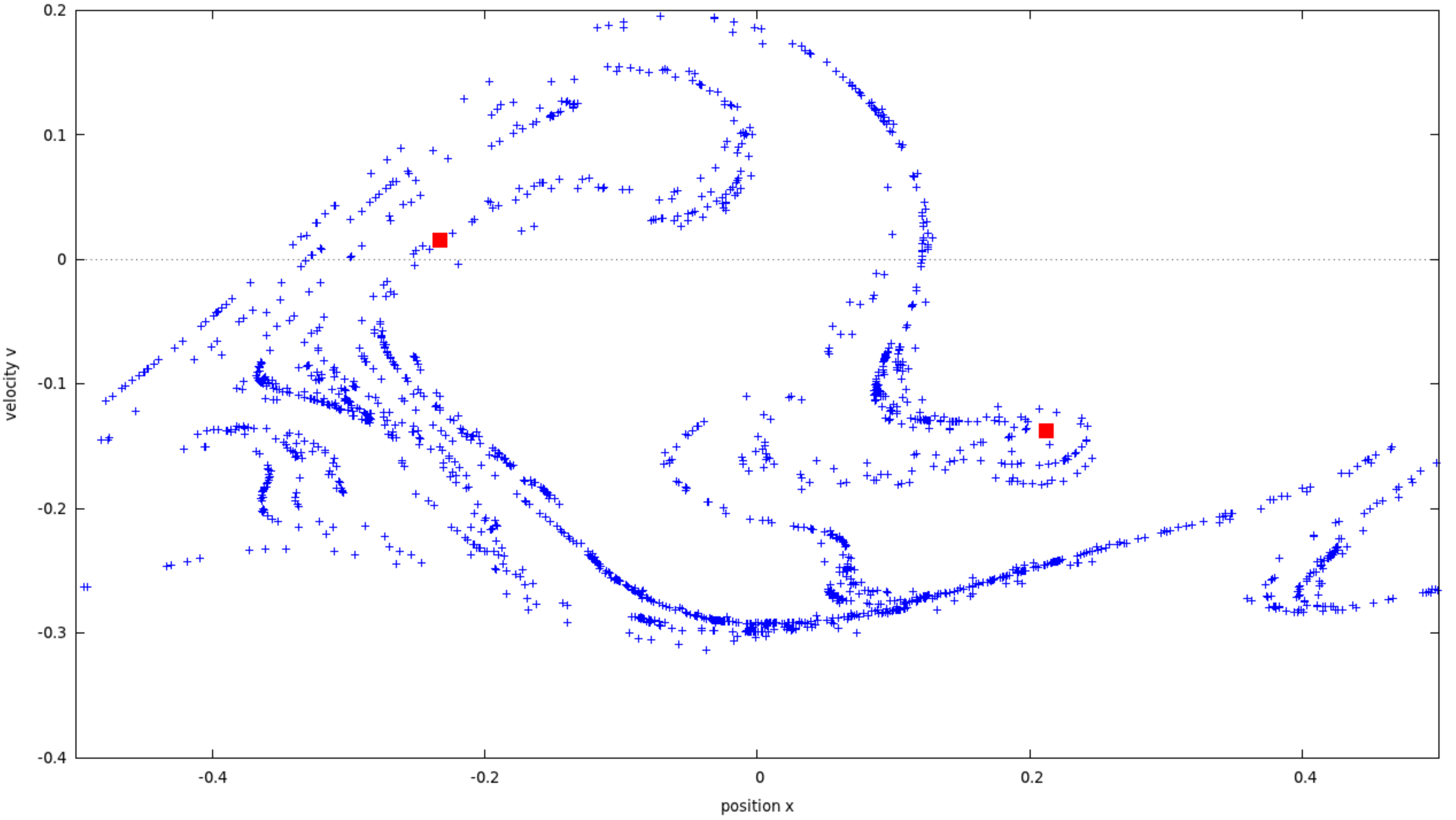}
   \caption{Poincare cross section for, for $Q=0.10304129010172704$, a typical Chaotic attractor (doted) and a period two attractor (square) plotted for for $\lambda=2.21$ and $\lambda=1.1$, respectively}.
	\label{fig2}
\end{figure}

To start out, let us recall that dynamics of our system based on the dimensionless Eq.~\ref{eq3} is essentially complex since chaotic attractors are predominant, thereby requiring a consistent statistical calculation; less to deviate expectations. Considering the current or the next flux of the system, a convergent recipes accounting for more elaborated statistical considerations has been deeply discussed and proposed~\cite{mateos2000,barbi2000,kenfack2007}. It was thus proven that a single trajectory is by far erroneous and that the fast and convergent one takes into account the transient as well as the large number of initial trajectories run. The current obtained from the net flux computed from a broad distribution\cite{jung1996,mateos2000,kenfack2007}, can be defined as:
\begin{eqnarray}
J=\frac{1}{N(M-n_c)}\sum_{j=n_c}^M\sum_{k=1}^Nv_k(t_j)
\end{eqnarray}  
as the time average of the average velocity over an ensemble of $N$ initial conditions randomly $x_k$ chosen in such a way to span the entire integration space of length $[-L,L]$, with arbitrary velocities $v_k(0)$. Here $v_k(t_j)$ is the velocity of a trajectory $k$ at a given observation time $t_j$, $M$ is the total observation times $t_j$ and $n_c$ is the transient time which is few hundreds of the period $T$. 
Because of the chaotic character of our system, bifurcation diagrams as well as Poincare cross sections, which we shall also consider computing in the subsequent section, are potential ingredients that may help to understand the transport properties of our system. Throughout the paper, the following parameters are kept constant: $b=0.10$ and $T=9.377888518179104$. Fig.~\ref{fig2} displays, for $Q=0.10304129010172704$, a prototypical chaotic attractor (doted) and period-two attractor (square) plotted in a Poincare cross-section for $\lambda=2.1$) and $\lambda=1.1$, respectively. What makes our system more complex is the fact there is a high probability of the coexistence of such attractors likely to transport in a unpredictable direction. 
\section{Results and discussions}
In this section we present results of our investigation focusing mainly on the computation of currents with associated bifurcation diagrams, Poincare cross sections and another type of current diagram newly introduced in this framework and which we conveniently term {\it current reversal diagram}. These are essentially orchestrated by the driving force which we purposely design here to have a full control over the chaotic transport of the system; in addition to the amplitude of the driving $Q$ commonly used in this field as control parameter, another parameter $\lambda$ acting as the driving phase has also been proposed (see Fig.~\ref{fig1} and Eq.~\ref{dforce}). Because $\lambda$ is responsible of the asymmetry of the driving force, it is called asymmetry parameter. 

\subsection{a) Influence of the asymmetric parameter $\lambda$}
\begin{figure}[hc!]
\includegraphics[width=0.85\linewidth,height=.35\textheight]{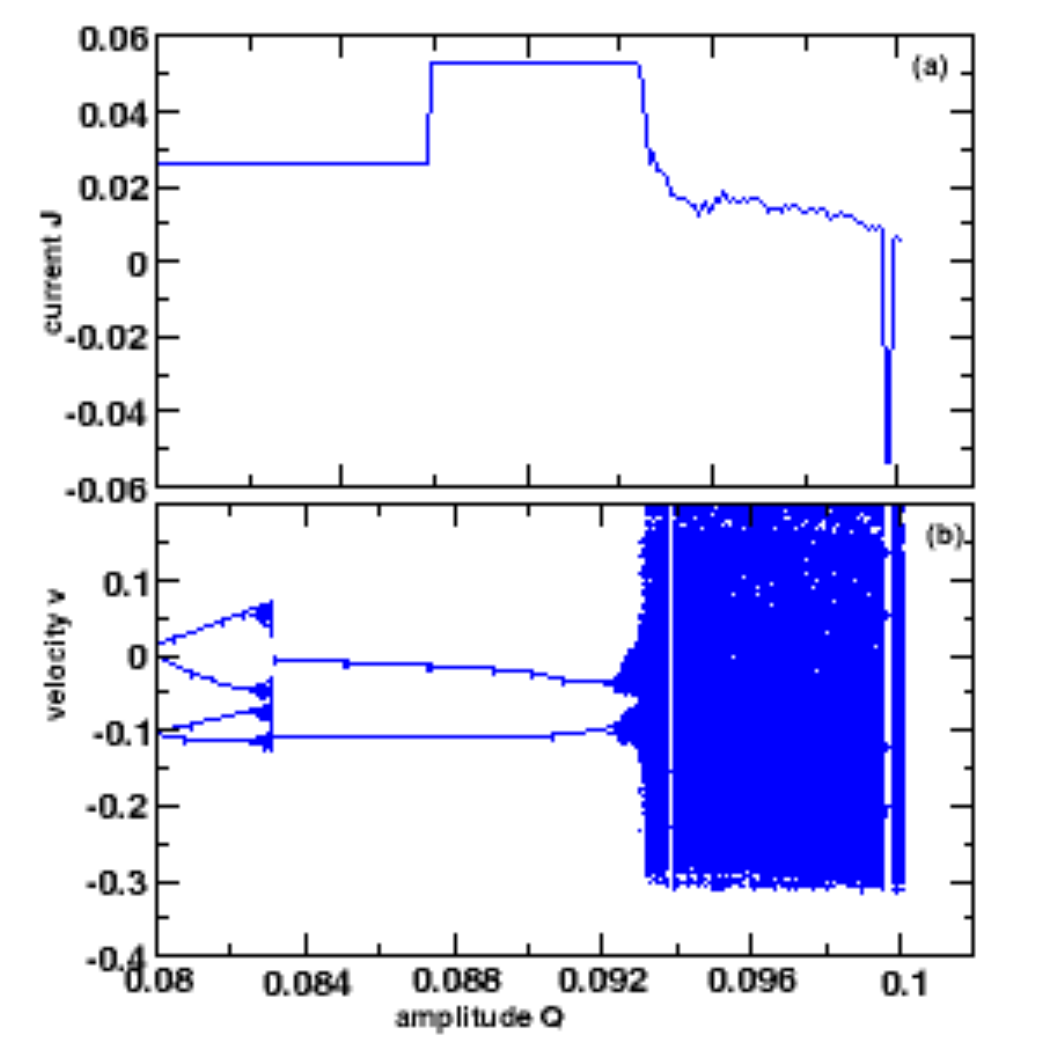}
   \caption{Current $J$ (a) and associated bifurcation diagram (b) plotted for a given value of $\lambda=2.52487$ and as function of the amplitude of the driving force $Q$. Note the behaviour of the current with spikes and specially the appearance of current reversal at around $Q=0.10$ clearly underlined with the transition periodic-chaos.}
	\label{fig3}
\end{figure}

\begin{figure}[hc]
\includegraphics[width=0.85\linewidth,height=.35\textheight]{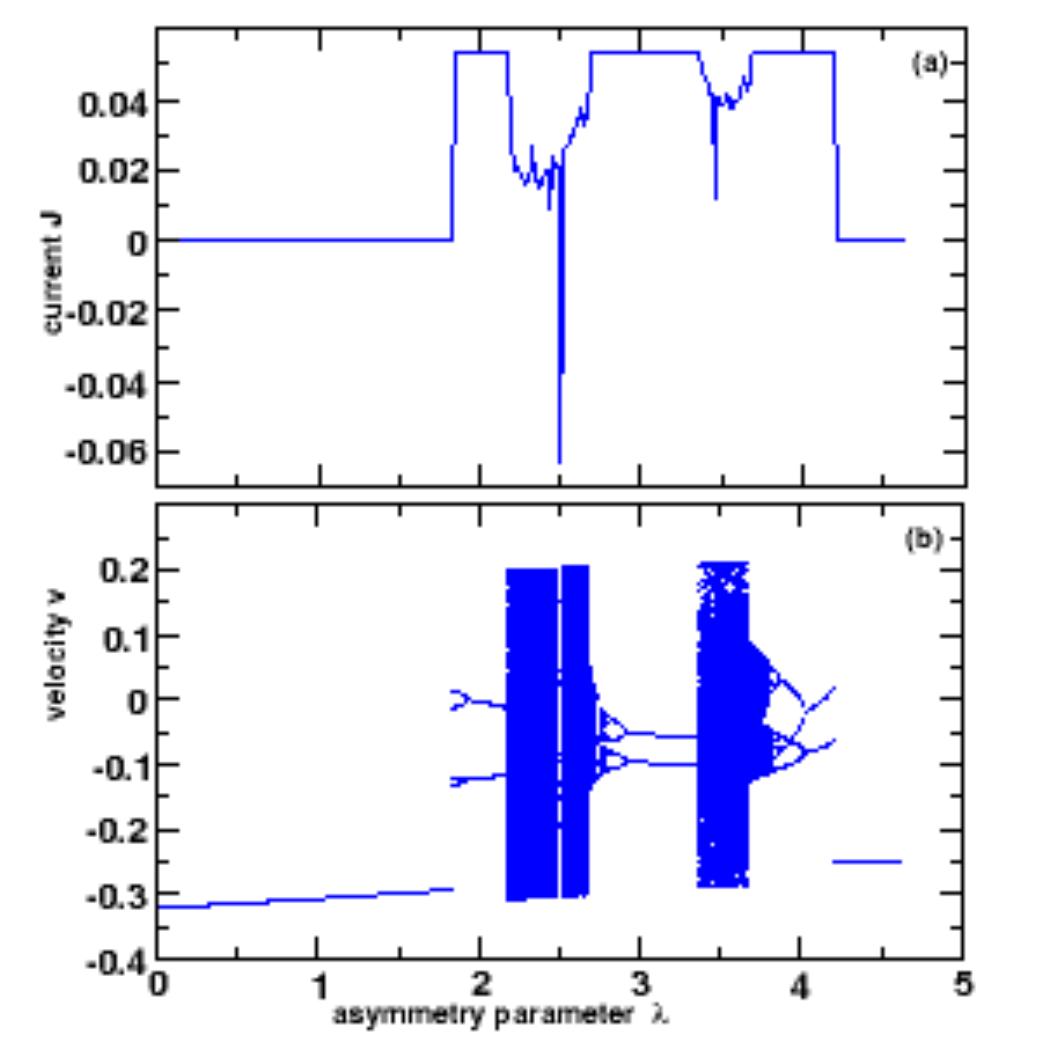}
   \caption{Current $J$ (a) and associated bifurcation diagram (b) plotted for a given value of $Q=0.09421972$ and as function of the asymmetric parameter $\lambda$. Like in Fig.~\ref{fig3} the appearance of reversals or jumps goes hand in hand with chaos. It turns out that just by changing $\lambda$ we can also drive the system to either he directed transport or to the reverse transport or current reversal.}
	\label{fig4}
\end{figure}

\begin{figure}[hc]
\includegraphics[width=0.85\linewidth,height=.35\textheight]{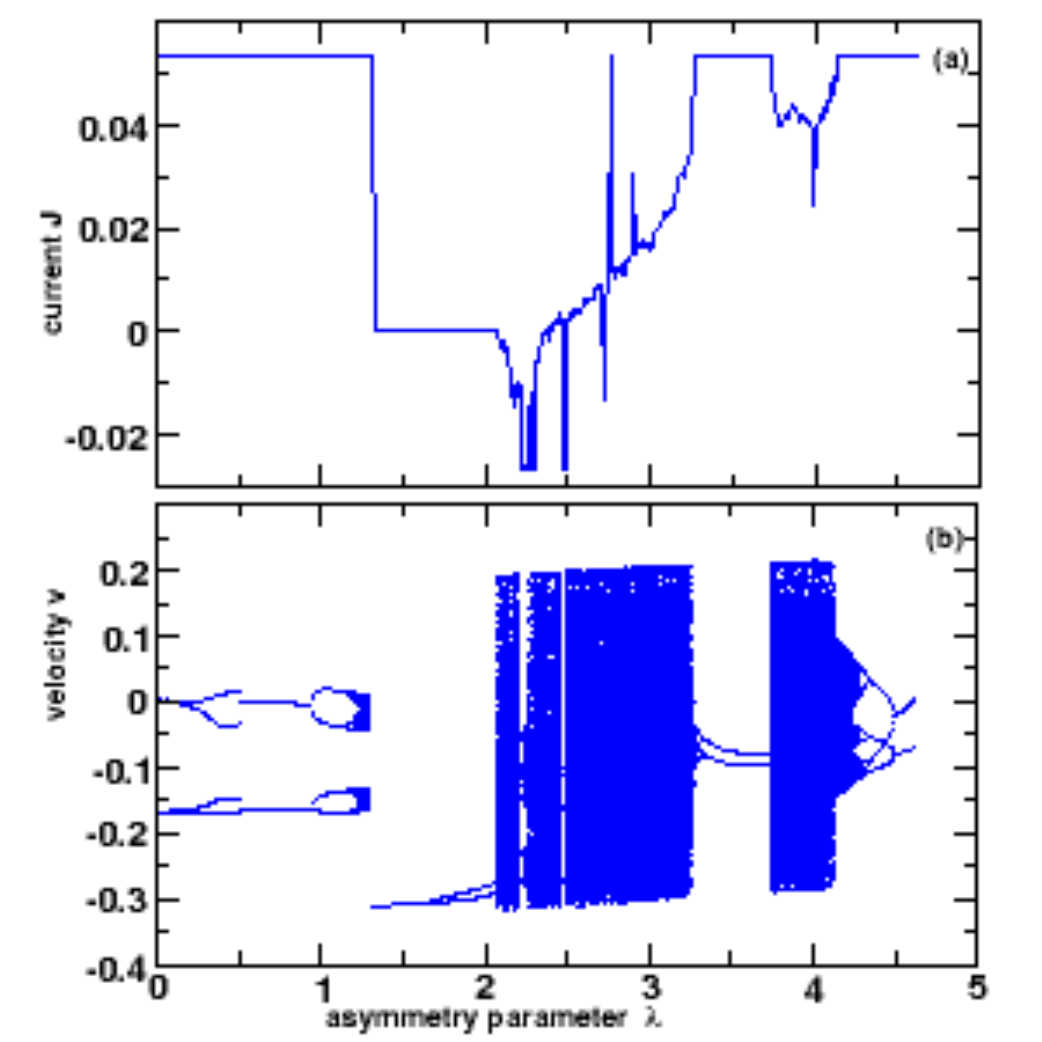}
   \caption{Same as in Fig.~\ref{fig4} but with $Q=0.10304129$ which induces a completely different sequence of current transport.}
	\label{fig5}
\end{figure}

In recent works, the phenomena of directed transport, current reversals and chaos have been largely addressed as function of the amplitude symmetry driving force. Of great importance, the origin of the counter intuitive current reversal (sudden reverse transport) has been demonstrated as associated with the chaotic background of the system. Here we want to see to which extent we can control such a non trivial transport system. To start out, we first plot in Fig.~\ref{fig3}, as function the amplitude $Q$ of the driving force, the current as well as its associated bifurcation diagram for $\lambda=2.52487$. As expected, the spikes observed in the  current transport $J$ (Fig.~\ref{fig3}-a) is guided by the choaticity of he system as illustrated by the bifurcation diagram (Fig.~\ref{fig3}-b). Remarkably, current reversal occurs at around $Q=0.10$, corresponding clearly to the transition chaos-periodic motions.
The question one may now raise up is whether one can also rectify current by means of the the asymmetric parameter $\lambda$ newly introduced as second control parameter. In other words, can one make use of $\lambda$ to induce a current reversal (which perhaps may not has no chaos connection) or to rectify an existing current reversal of the system? In what follows we focus our attention to the behaviour of the current transport as function of the asymmetry parameter ($\lambda \in (0, 4.5]$)for few arbitrary values of the amplitude $Q$. Fig.~\ref{fig4}, plotted for $Q=0.09421972$, clearly exhibits current intermittency (Fig.~\ref{fig4}-a) corresponding to regions of directed transport, reverse transport and non transport ($J=0$). This behaviour along the asymmetry parameter $\lambda$ is one to one mapped with the associated bifurcation diagram (Fig.~\ref{fig4}-b), thus demonstrating that these features go hand in hand with chaos present in the system. With a slightly higher value of amplitude, say $Q=0.10304129$, a completely different sequence of current opportunities shows up (Fig.~\ref{fig5}), again well supported with the corresponding bifurcation diagram. Along these lines we have tested several other values of $Q$ (no shown) and the resulting trend is qualitatively the same, thus showing how rich is our system. These results are clear indications that the asymmetry parameter can also assist in controlling the current reversals and suggest the full control over the two parameters ($Q,\lambda$) which we intend to investigate subsequently. 

\subsection{b) Two parameters current reversal diagram}
\begin{figure}[hc]
\includegraphics[width=0.55\linewidth,height=.25\textheight]{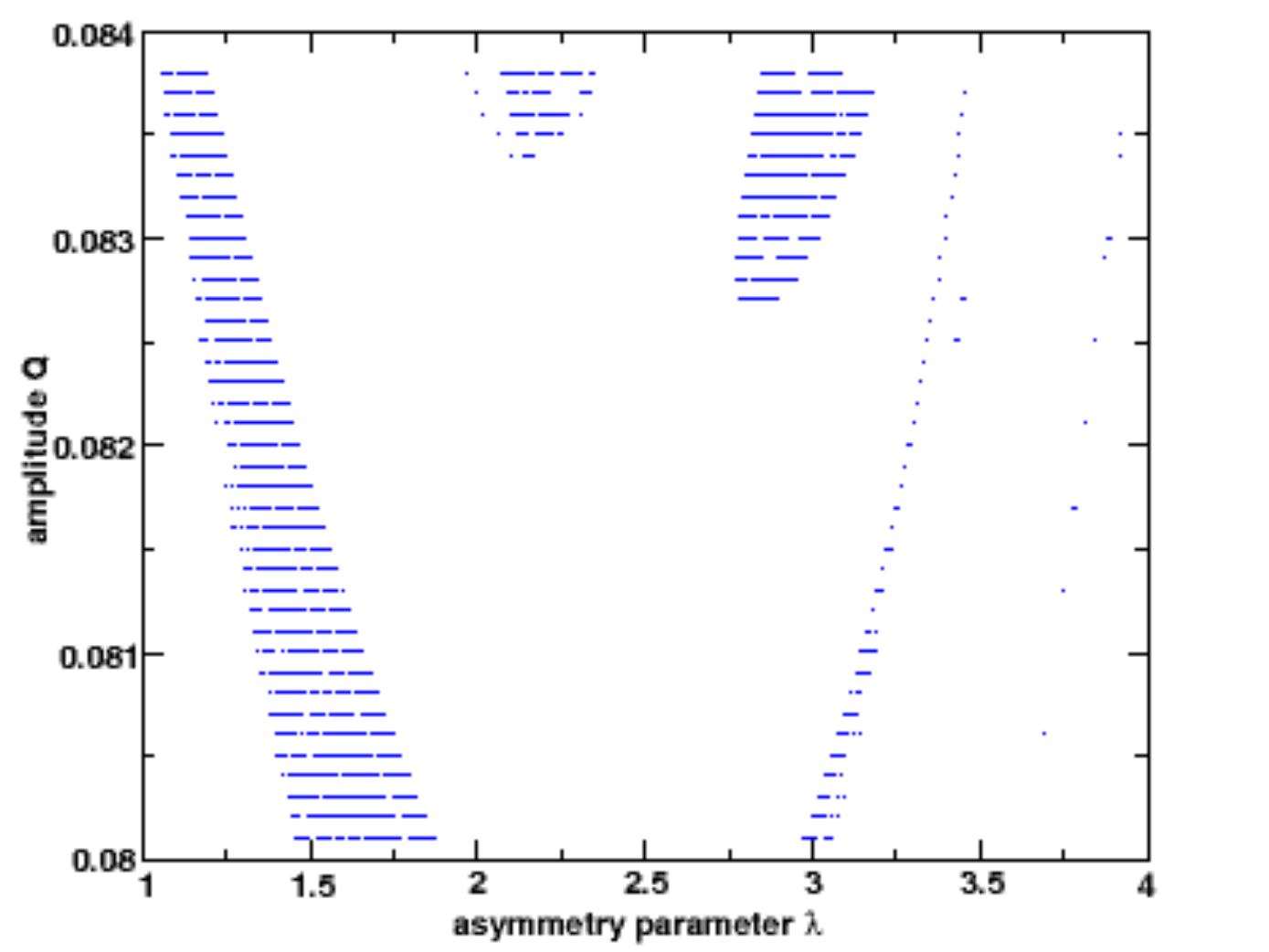}
   \caption{Current reversal diagram as function of the amplitude $Q$ and the asymmetric parameter $\lambda$. Regions of current reversals (shaded) and no current reversal or no current (white) are clearly shown.}
	\label{fig6}
\end{figure}

We have demonstrated above that the current reversal strongly depends on the amplitude $Q$ of the driving force and the asymmetry parameter. For a more elaborated and general viewpoint, a current reversal diagram has been carefully computed and plotted in Fig.~\ref{fig6} as function of $Q$ and $\lambda$. Regions of current reversals (shaded) as well as regions of no current reversals (white) can clearly be seen, thereby providing us with a reliable spectrum of the system. By choosing a couple of control parameters ($Q,\lambda$) we fall of course in one or another transport regime. This figure is expected to be of importance for experimentalists working in this field.

\section{Conclusion}
To conclude we have revisited the one dimensional deterministic chaotic ratchet model, now replacing the driving force with an alternative purposely tailored to allow for the full control of a particle transport of the system.  The asymmetry of this external force newly introduced here plays a key role as the direction of the transport can significantly be altered by changing the asymmetry parameter $\lambda$. Our results thus lead to the establishment of conditions for observing zero transport, directed transport and reverse transport or current reversal. By means of a current reversal diagram depending on two control parameters, namely the amplitude $Q$ and the asymmetry parameter $\lambda$, we have been able to provide a more general spectrum exhibiting a high visibility of current transport in the system. This way of fully exploring the appearance of current reversal here is expected to be of great interest in experiments.


\begin{theacknowledgments}
This work has been supported by the African University of Science and Technology (AUST) in Abuja.
AK is very grateful to AUST for hosting him during his recent research
and teaching visits which have led to this collaboration. DSB and OA-O
acknowledge the Centre of High-Performance Computing (CHPC) in South
Africa and the Abdus Salam International Centre for Theoretical Physics
(ICTP) in Trieste, Italy for computational resources.
\end{theacknowledgments}



\bibliographystyle{aipproc}   


 {\typeout{}
  \typeout{******************************************}
  \typeout{** Please run "bibtex \jobname" to optain}
  \typeout{** the bibliography and then re-run LaTeX}
  \typeout{** twice to fix the references!}
  \typeout{******************************************}
  \typeout{}
 }

\end{document}